\documentstyle[12pt]{article}
\begin{document}
\vskip 2 cm
\begin{center}
\Large{\bf QUANTUM ENTANGLEMENT AND QUANTUM CHROMODYNAMICS } 
\end{center}
\vskip 2 cm 
\begin{center}
{\bf AFSAR ABBAS}
\vskip 3 mm
Institute of Physics, Bhubaneswar-751005, India

(e-mail : afsar@iopb.res.in)
\end{center}
\vskip 20 mm
\begin{centerline}
{\bf Abstract }
\end{centerline}
\vskip 3 mm

Non-locality or entanglement is an experimentally well established
property of quantum mechanics. Here we study the role of quantum
entanglement for higher symmetry group like $ SU(3_c) $,
the gauge group of quantum chromodynamics ( QCD ). We show that the
hitherto unexplained  property of confinement in QCD arises as a
fundamental feature of quantum entanglement in $ SU(3_c) $.

\newpage

Quantum entanglement, leading to non-locality, is an experimentally well
tested aspect of quantum mechanics [1,2,3,4,5,6]. Starting with two
particles, now four particle entanglement has been demonstrated [7].
Non-local quantum entanglement forms the basis of the concept of quantum
information thereby enabling such phenomena as quantum cryptography [8],
dense coding [9], teleportation [9] and quantum computation [10].

The experimental demonstration of quantum entanglement involves 
polarization states of the photons, spin of the electrons and atoms
etc. Hence objects within $U(1)_{em}$ and representations of the spin
SU(2) group are involved. 
This entanglement necessarily implies non-locality.
Here we would like to study the property of quantum entanglement
for representations of higher group, in particular the group SU(3).
Note that the group $SU(3_c)$ ( here the subscript c is placed to
indicate color degree of freedom ) forms the basis of the theory
of strong interaction, the Quantum Chromodynamics, While the 
electrons and photons are immune to QCD, atoms for whom the quantum
entanglement and non-locality has also been demonstrated,
are composite systems - consisting of electrons and a nucleus.
A nucleus is made up of protons and neutrons whose structure and 
properties are determined by the rules of QCD. An understanding 
of non-locality in the quantum atomic systems would involve an 
understanding of quantum entanglement in QCD. In fact viewed
in this manner, it becomes a puzzle as to why quantum non-locality
of elementary systems like the electron and the photon is the same as 
in the composite systems like the atoms. How does the quantum entanglement
function in the QCD case to make the above possible?
One should also not forget that after all the protons and the 
neutrons are themselves composite systems of quarks and gluons 
which in turn are governed by QCD. Hence this understanding 
may have implications for quantum measurements, quantum cryptography,
quantum teleportation and quantum computation.

As a theory of strong interaction QCD is well established both
theoretically and experimentally. This is well documented in 
literature. ( eg. see ref. [11] ). Herein, the protons and neutrons
are made up of three constituent quarks and the mesons are made up
of a quark-antiquark pair. Note that the quarks belong to the 
fundamental representation  3 of $SU(3_c)$, the gauge group of
QCD. In the colour space

\begin{equation}
3 \otimes 3 \otimes 3 = 1 \oplus 8 \oplus 8 \oplus 10
\end{equation}

\begin{equation}
3 \otimes\overline 3 = 1 \oplus 8
\end{equation}

The above coupled with the fact that single coloured quark or any other
coloured object like a gluon have not been found in nature, has led to
the hypothesis that colour is permanently confined and that colour
singlet objects are only found in nature. This so called colour singlet
confinement hypothesis is well tested experimentally. Right from the
early days of QCD there have been intense efforts to prove the
confinement hypothesis in QCD. There been intriguing hints of confinement
in some interpretations, however as of now there has been no definitive
proof of confinement in QCD. It is widely expected that confinement arises
due to some non-perturbative property of QCD. But, as to what it is,
nobody knows. Hence the problem of confinement is still an open one
in QCD.

When three quarks or a quark-antiquark come together only colour 
singlet representations are observed. Hence in the canonical picture
it is proposed that in the colour space 
while the singlet has a finite mass, the octet, 
decuptet ( and all other higher representations arising in multiquark
systems ) are infinitely heavy and hence do not contribute to the low
energy spectrum. Also the triplet representation ie. the free quark
becomes infinitely heavy when free. Hence the singlets have finite mass
and the coloured masses are infinitely heavy [11,12]. Thus it has been a
long standing puzzle as to how infinitely heavy free objects can come
together in a size of 1 fm to give finite mass bound states.

To appreciate the point further, let us go to the hidden colour concept
in the multiquark system. When a neutron and a 
proton come together to form a bound state of deuteron of approximate size
of 2 fm, then at the centre in a region less than a fm or so they should
overlap to appear like a six quark bag. As per the colour singlet
hypothesis the 6-quark bag looks like,
 
\begin{equation}
| 6q > = \frac{1}{ \sqrt{5} } |1>|1> + \frac{2}{ \sqrt{5} } |8>|8>
\end{equation}

where $|1>$  represents a 3-quark cluster which is singlet in colour space
and $|8>$ represents the same as octet in colour space. Hence $ |8>|8> $
is overall colour singlet. This part is called the hidden colour 
because as per confinement ideas of QCD
these octets cannot be separated out asymptotically and so manifest
themselves only within the 6-q colour-singlet system [13].
Group theoretically the author had earlier obtained the hidden colour
components in 9- and 12-quark systems [14,15].  
The author found that the hidden colour component of the 9-q
system is $ 97.6 \% $ while the 12-q
system is $ 99.8 \% $ ie. practically all coloured.
These 9- and 12-quark configurations have been found to be relevant in
nuclear physics for the A=3,4 nuclei 
$ ^{3}H $, $ ^{3}He $ and $ ^{4}He $. 

Now the problem is that three infinitely heavy quarks come together to
form a finite mass colour singlet object while at the same time forming 
infinite mass octet and  decuptet states. When two such colour singlet
objects as in eq. 3 come together both the individual singlet as well
as octet parts contribute. How come 6-q singlet arises due to two finite 
mass colour singlets and two infinite mass colour octet objects, both of
which contribute similar amounts to the total mass of the 6-q system?
Similar problems persist for the 9-q and the 12-q systems [14,15].

Note that what we are doing here is identical to the configuration
mixing problem in quantum mechanics - in particular nuclear physics.
To determine various properties in nuclear physics one finds that one has
to mix several states with a particular quantum number to obtain 
agreement with the experiments. The configurations  
which mix to give a particular state are all degenerate or very close in
energy. It is always found that configurations which are separated
by large energy gaps do not contribute significantly to to a particular
state.
In principle one has an infinite dimensional Hilbert space, however only
a finite number of configurations are known to mix to give a state of
good quantum number which is physically relevant. 
Hence in quantum mechanics if any configurations
are infinite apart in energy then they are infinitely suppressed in
the mixed final state. Therefore it is a puzzle here is as to how the two
3-q states configurations the singlet and the octet, 
which are infinite energy apart in energy
manage to mix in the 6-q system. However this is precisely what
confinement means for the 6-q system. Now looking back at eqns 1 and 2
we notice that for confinement in the case of baryons and mesons 
the same problem of configuration mixing
arises. So how does this come about? Thus the confinement problem should
be viewed as a fundamental outstanding problem in quantum mechanics.

As shown above, for the ground state hadrons at 
temperature T=0, the singlet states have finite energies and the coloured
states 3, 8, 10, 27 etc are all expelled to infinite energies.
So far there has  been no clue as to why it is so in QCD, though there
have been several model calculations indicating this feature [11,12].
Let us now leave the
T=0 region and proceed to some finite temperature region and study the 
problem there. Though it has never been explicitly demonstrated 
even in a toy model calculation, there is a common feeling that at finite
temperatures too the same infinite
separation between the singlet and the coloured objects would persist. 

Recently we have looked at this specific problem [16].
We looked at 
the role of higher representations like 8-plet, 27-plet etc.
for large hadronic systems like quark stars and objects
created in high energy heavy ion collisions.

The orthogonality relation for the associated characters 
$ \chi_{(p,q)} $ of the (p,q) multiplet of the group $ SU(3)_c $
with the measure function $ \zeta(\phi,\psi) $ is [16]

\begin{equation}
\int_{SU_{C}(3)}^{} d \phi \, d \psi \, \zeta \left( \phi,\psi \right) 
\chi^{\star}_{(p,q)}\left( \phi,\psi \right) 
\chi_{(p',q')} \left( \phi,\psi \right) = \delta_{pp'} \delta_{qq'}
\end{equation}

Let us now introduce the generating function
$ Z^{G} $ as

\begin{equation}
Z^{G}(T,V,\phi,\psi) = \sum_{p,q} \frac{ Z_{(p,q)} }{ d(p,q) }
\chi_{(p,q)} ( \phi,\psi )  
\end{equation}
with

\begin{equation}
Z_{(p,q)}=tr_{(p,q)} \left[ 
\exp{ \left( -\beta \hat{H}_{0} \right) } \right]
\end{equation}

$ Z_{(p,q)} $
is the canonical partition function.
The many-particle-states which belong to a given multiplet (p,q)
are used in the statistical trace 
with the 
free hamiltonian $\hat{H}_0$, d(p,q) is
its dimensionality and $\beta$ is the inverse of the temperature T.
The projected partition function $ Z_{(p,q)} $
can be obtained by using the orthogonality relation for the 
characters. Hence the projected partition function for any representation
(p,q) is

\begin{equation}
Z_{(p,q)}= d(p,q)
\int_{SU(3)_{c}} d\phi \, d\psi \, \zeta \left( \phi,\psi \right) 
\chi^{\star}_{(p,q)}(\phi,\psi)
Z^G \left( T,V,\phi,\psi \right)
\end{equation}

Once we have the partition function for any 
representation  $Z_{(p, \ q)}$, 
then any thermodynamical quantity of interest can be calculated. 
For example the energy 
\begin{equation}
 E_{(p, \ q)} \ = \ T^2 \frac{\partial}{\partial T} \ln Z_{ (p, \ q) } .
\end{equation}

In ref. [16] we projected out different representations like singlet
(0,0),
octet (1, 1), 27-plet (2, 2) etc. on these large hadrons. The most 
interesting result  we obtained is that for large values 
of $TV^{1/3}$ ( where V is a measure of the size of the composite object ) 
all representations ; singlet, octet, 27-plet etc are all
degenerate in energy ( ie they all have the same energy ) 
with the unprojected state. There is nothing which favours the
colour-singlet 
representation over the colour-octet at high temperatures. 
At the same time we also found  that our
projection technique at low temperatures is able to discriminate between
the singlet and the octet states etc by clearly favouring the singlet
state over others which are all expelled to infinite energies..

This result is quite general and at high enough temperatures represents
the generic property of QCD that all representations  have the same energy
[16]. Only as the temperature drops do the singlets find themselves
favoured over the other colour states and all the coloured states are
expelled to infinite energies. 

What is our singlet state at finite
temperature? It is degenerate in energy with respect to all coloured
states. This colour singlet is made from configurations like
$1 \otimes 1$, $8 \otimes 8$, $27 \otimes 27$ etc, that is, all
permitted coloured states which can come together to give a singlet.
Note that in the said paper [16] we had looked at the chemical potential
zero state. We have found that the same holds for non-zero chemical
potential as well. Since all these have the same energy, as per the
configuration mixing idea in quantum mechanics, at finite temperature 
there is nothing barring them from contributing to the colour
singlet object. How this significant effect at finite temperature
manifests itself at T=0 is what we shall discuss below.

As per the standard picture of cosmology the Universe was much
hotter at early times. Hence at high enough temperature in the Early
Universe as per our calculation in QCD [16], all the coloured states must
have existed and were degenerate with the colour singlet
states. Being degenerate in energy all kind of configuration mixing giving
a particular quantum states were allowed quantum mechanically.
Hence, the 6-q state as given by eq. 3 was a quantum mechanically
permitted configuration mixed state. In this particular case in the Early
Universe clearly the mixing configurations were the singlet and the octet.
It was at this stage in the Early Universe that the two states, $|1>$ 
with another $|1>$ and $|8>$ with another $|8>$ got entangled
quantum mechanically. Similar things happened for all other possible
states. Hence a 3-q colour singlet state was an entangled state of three
different $|3>$ coloured states at that temperature. And so on.

As the Universe cooled as shown by us [16], the colour singlet states
would come down in energy while the coloured states would be
expelled to infinite energies. 
And this happens to be the physical situation at present. 
Hence today when the $|6-q>$ state has components $|1>|1>$ 
and $|8>|8>$, it is because the
system remembers that it was entangled in this manner in the Early
Universe. It is a manifestation of quantum entanglement of these
states in the Early Universe which manifests itself
as confinement in QCD at present. We can argue in the same manner for 
all possible colour singlet states available at present.
Hence we propose here that confinement in QCD at present arises 
as a result of quantum entanglement in $SU(3_c)$ at finite temperature
in the Early Universe.

Note that quantum entanglement in the case of the spin properties of
electrons and photon manifests itself as non-locality. In the case of QCD,
as shown here, quantum entanglement appears as the local colour
confinement. It is quantum correlation which is important. Non-locality or
locality depends upon the quantum property under consideration,

Note that the significant effect of the quantum entanglement in QCD 
leading to confinement is that this is a local effect, meaning that 
it is ensures that QCD entanglement acts only within a distance of a 
fermi for a nucleon and a few fermis for a nucleus.
This ensures that whether you use an electron ( or a photon ) beam,
the quantum entanglement effects as studied in the two slit experiments
or the EPR kind of experiments, the effect would be the same if you used
an atomic beam instead. Had the quantum entanglement effect in QCD been
any different than to give confinement in a finite size, the effects
for the Universe today would have been unimaginably different!
Actually this fact should be seen as experimental confirmation of the
ideas presented here.

In summary, we emphasize our recent result in the study of QCD at 
finite temperature, that all states - coloured as well as singlet are 
degenerate at finite temperatures [16]. This allows for all kind of
quantum configuration mixing giving a particular state
in the Early Universe. Hence these 
configurations get quantum mechanically entangled. As the Universe 
cooled to reach the present status of finite mass colour singlet states
and infinite mass coloured states, the colour singlet states remember as
to how they were made up of which coloured entangled states 
when the Universe was hot. And continues to behave that way at present.
Hence confinement in QCD at present, arises as a result of quantum
entanglement in QCD at finite temperature in the Early Universe.

\newpage 
\vskip 5 mm
{\bf References}
\vskip 5 mm
1. J F Clauser and A Shimony, Rep Prog Phys {\bf 41} (1978) 1881

2. A Aspect, P Grangier and G Roger, Phys Rev Lett {\bf 49}
(1982) 91

3. W Tittel, J Brendel, H Zbinden and N Gisin, Phys Rev Lett
{\bf 81} ( 1998) 3563

4. G Weihs, T Jennewein, C Sunon, H Weinfurter and A Zeilinger,
Phys Rev Lett {\bf 81} ( 1998) 5039

5. P G Kwiat, E Waks, A G White, I Appelbaum and P H Eberhard,
Phys Rev {\bf A 60} ( 1999) R773

6. S Duerr, T Nonn and G Rempe, Nature {\bf 395} (1998) 23

7. C A Sackett, D Kielpinski, B E King, C Langer, V Meyer, C J Myatt,
M Rowe, Q A Turchette, W M Itano, D J Wineleand and C Monroe,
Nature {\bf 404} (2000) 256

8. A K Ekert, Phys Rev Lett {\bf 69} (1992) 1293

9. C Bennett et al, Phys Rev Lett {\bf 70} (1993) 1895

10. Ed.- D DiVencenzo, E Knill, R LaFamme and W Zurek,
Proc R Soc Lon Ser {\bf A 1969} (1998)

11. R E Marshak, " Conceptual foundations of modern particle 
physics", World Scientific Pub Ltd ( Singapore ) 1993

12. C Bender, Phys Rep {\bf 75} (1981) 205

13. V A Matveev and P Sorba, Lett Nuovo Cim {\bf 20} (1977) 435

14. Afsar Abbas, Phys Lett {\bf B 167} (1986) 150

15. Afsar Abbas, Prog Part Nucl Phys {\bf 20} (1988) 181

16. Afsar Abbas,Lina Paria and Samar Abbas, Eur Phys J 
{\bf C 14}(2000)695 \\
Journal electronic version: http://dx.doi.org/10.1007/s100520000395

\end{document}